\documentclass[prd,twocolumn,draft]{revtex4}
\usepackage{bm}
\usepackage{hyperref}
\usepackage{latexsym}
\usepackage{mathrsfs}
\usepackage{graphicx}
\usepackage{graphics}
\usepackage{txfonts}
\usepackage{fancybox}
\usepackage{helvet}
\usepackage{fancyhdr}
\input epsf.sty
\font\titlefont=cmssbx18 
\font\subtitlefont=cmssbx10
\font\authorfont=cmr12
\font\abstractfont=cmr10

\font\sectionfont=cmssbx10
\font\subsectionfont=cmssbx10

\font\figtextfont=cmss9

\font\reftitlefont=cmssbx10 at 12pt
\font\reftextfont=cmr10

\begin{document}
\renewcommand{\theequation}{\arabic{equation}}
\title{{\titlefont The Twin Paradox Revisited and Reformulated}\\ {\subtitlefont On the Possibility o\textit{f} Detecting Absolute Motion }}

\author{\authorfont G. G. Nyambuya${}^{1}$}
\email{gadzirai@gmail.com, fskggn@puk.ac.za}
\author{\authorfont M. D. Ngobeni${}^{1,2}$}
\email{diengovza@gmail.com, fskmdn@puk.ac.za}

\affiliation{%
\authorfont \small \vspace*{0.5cm}${}^{1}$ North-West University -Potchefstroom Campus, School of Physics - Unit for Space Research, Private Bag X6001, Potchefstroom, Republic of South Africa.\\ ${}^{2}$Vaal University of Technology, Faculty of Engineering Department of None-Destructive Testing, P. Bag X021, Vanderbijlpark, 1900, Republic of South Africa.} 

\date{\today}

\date{\today}

\begin{abstract}
\begin{center}

\end{center}
\linethickness{2pt}
\line(1,0){400}\\
\\
$\textbf{\Large {\sectionfont Abstract.}}$ \abstractfont The famous twin paradox of the Special Theory o\textit{f} Relativity by Einstein (1905) is revisited and revised. This paradox is not a paradox in the true sense of a paradox but a reflection of a misunderstanding of the problem and the Principle o\textit{f} Relativity. The currently accepted solution to this takes into account the accelerations and deceleration of the traveling twin thus introducing an asymmetry that solves the paradox. We argue here that, with the acceleration and deceleration neglected, the problem is asymmetric hence leading to the same conclusion that the traveling twin will age less than the stay at home. We introduce a symmetric twin paradox whose solution can not be found within the currently accepted provinces of the STR if one adopts the currently accepted philosophy of the STR namely that it is impossible for an inertial observer to determine their state of motion. To resolve this, we present (in our modest view) a simple and convincing argument that leads us to conclude that it must be possible for an inertial observer to determine their own state of motion. With this, we are able to solve the symmetric twin paradox. The fact that it is possible for an inertial observer to determine their state of motion -- brings us back to the long rejected idea of an all pervading and permeating medium -- the Aether, namely the Lorentz luminiferous Aether. An experiment capable of validating or invalidating this claim is suggested.
\line(1,0){400}

\textsl{\begin{center}
``There is no absolute space, and we only conceive of relative motion;\\ and yet in most cases mechanical facts are enunciated as if there\\ is an absolute space to which they can be referred.''
\end{center}}

\begin{flushright}
-- \textbf{Jules Henri Poincar\'e} (1854-1912)
\end{flushright}
\end{abstract}

\maketitle

\section{\sectionfont Introduction}

The philosophy derived from the Principle o\textit{f} Relativity, according to which the Laws o\textit{f} Physical Phenomena must be the same for a ``stationary'' inertial observer as for one that is in uniform relative motion with the ``stationary'' inertial observer, states that there exists no means by which any inertial observer can determine whether or not they are in motion.
This philosophy introduces some uncomfortable inconsistencies that have made some critics  of the STR to spend a considerable amount of their time (such as Professor Herbert Dingle who spent about thirty years, see e.g. McCausland 2008) arguing that these inconsistencies rendered the STR obsolete. The STR has never failed any experimental test to which has been subjected to, and this has lead to the mainstream scientific community to ignore any such criticism.

This philosophy that there exists no means -- mechanical or optical -- by which any inertial observer can determine whether or not they are in motion rests its weight on the Michelson-Mosley Experiment (MM-Exp) (Michelson 1881, 1887). The MM-Exp is an experiment that was designed to measure the speed of the Earth in the hypothetical Aether. This Aether was thought to exist since James Clerk Maxwell had shown that light was a wave and this light wave travelled at a constant speed denoted by the symbol $c$. Since a wave needs a medium which to travel in; it was supposed that the Aether filled the whole Universe and was an absolute stationary frame of reference which was rigid to electromagnetic waves but completely permeable to ponderable matter. The MM-Exp was then designed and conducted. Much to suprise of the then present scientific community, the experiment showed no proof of the existence Aether or lack thereof. Without the knowledge of the MM-Exp, Albert Einstein reasoned  that it was not necessary to invoke this hypothetical medium. He reasoned, that, if the Laws o\textit{f} Physical Phenomena where to be the same for all inertial observers, and  speed of light where an absolute constant as predicted by Maxwell's theory, then the speed of light ought to be a Universal and absolute constant for every observer in the Universe everywhere and everytime. We argue here in favor of the Aether and we here cajole the reader to not stop but proceed. Normally, well seasoned physics readers couldn't give much time to a reading that tries to revive the Aether hypothesis -- hence the cajoling effort to draw the reader.

As aforesaid, this reading seeks to revive the Aether hypothesis -- we call upon the dead to arise and shine. We re-examine closely the long held underlaying philosophy of the STR emanating from the Principle o\textit{f} Relativity via the twin paradox gedanken. First we give an exposition of the well known twin paradox whereafter a modified version of is propounded. This modified version is -- unlike the the original version, symmetric in every respect. The symmetric nature of the new version, brings about an inconsistency that the STR is unable to solve. Even if the General Theory o\textit{f} Relativity (GTR) where to be brought to the STR's rescue, this inconsistency is insoluble unless we critically and carefully revise the underpinning philosophy emanating from the Principle o\textit{f} Relativity -- namely that: 

\textsl{\textbf{``No inertial observer can determine their state of motion by any means -- be it via mechanical means, optical means or otherwise; it is simple impossible.''}} 

We persuade and ask the reader to approach this reading with an open mind - that is, devoid of the now seemingly entrenched philosophy that Einstein's STR trashed once and forever the Aether to the ``Science Museum \textit{of} Great \textit{but} Failed Ideas''. 

The idea of the possibility that absolute motion may be real has support from other some authors see for example Cahill (2007, 2008). Cahill approaches this topic from a different vantage point known as process physics which seems to have some experiment basis. The detection of absolute motion directly leads to the Aether hypothesis and the works Cahill (2007, 2008) surely point to this.

\subsection{\subsectionfont Twin Paradox (Asymmetric)}

When it comes to the STR, a natural source of confusion for those encountering the STR for the first in their endeavor to comprehend the time-dilation effect is summed up in the so-called twin paradox, which is not really a paradox in the true sense of a paradox. This so-called paradox goes as follows: Suppose we have a set of twins, Taurai and Taurwi, and Taurwi decides to celebrate his $21^{th}$ birthday in style by rocketing at a constant relativistic speed (i.e. speeds comparable to the speed of light, for which the effects predicted by the STR become important and easily measurable) to the nearest star to planet Earth -- Alpha-Centauri which is $4.26$ lightyears away and this is incordance with a ``stationery'' Earth bound observer. Taurai and Taurwi are recent physics graduates who understand very well Albert Einstein's (1905) STR. Taurwi makes a round-trip, that is, he travels to Alpha-Centauri at a constant speed and upon arriving he returns to mother Earth. The other twin Taurai, decides to stay at home and not join his adventurous twin brother.

According the STR, Taurai sees Taurwi as moving away from the Earth and at the sametime, Taurwi has equal claim in his own frame of reference that he is not moving but Tauri is, that is he [Tauri] is moving away from him at the same speed as that Taurai sees him move albeit in the opposite direction. The paradox arises because according to the STR, the one that is moving will experience time dilation -- so the question is; since each sees the other as moving, then, who amongst them will experience this time dilation? and thus seem younger to the other.

This apparent paradox arises from an incorrect application of the Principle o\textit{f} Relativity to the description of the story by the traveling twin's point of view and the widely accepted resolution of this apparent paradox goes as follows. From his [Taurwi] point of view, the argument goes; his non-adventurous stay-at-home brother is the one who travels backward on the receding Earth, and then returns as the Earth approaches the spaceship again, while in the frame of reference fixed to the spaceship, the astronaut twin is not moving at all. It would then seem that the twin on Earth is the one whose biological clock should tick more slowly, not the one on the spaceship. 

The ``flaw'' in the reasoning is that the Principle o\textit{f} Relativity only applies to frames that are in motion at constant velocity relative to one another, i.e., inertial frames of reference. The astronaut twin's frame of reference, is a noninertial system, because his spaceship must accelerate when it leaves, decelerate when it reaches its destination, and then repeat the whole process again on the way back home. Their experiences are not equivalent, because the astronaut twin feels accelerations and decelerations thus leading to the conclusion that the traveling twin  will be younger when they are reunited. Given this solution and that this typically presented in books that deal with the STR at length, it is suprising that some notable authors (see e.g. Kark 2007) still regard the twin paradox as a paradox.

The real trick is the accelerations and decelerations experienced by the traveling twin; these  bring about the asymmetry which leads to Taurwi being the one that experiences the time dilation. From the purely idealized point of view,  we can neglected these accelerations and decelerations. If we did this -- will the scenario be symmetric? Since it is these accelerations and decelerations that bring in the asymmetry, it must follow that we must have a paradox because symmetry ought in this case to be restored thus leading to a real paradox. 

We wish to point-out here that our treatment of the twin paradox since it was conceived has been erroneous because the idealized twin paradox without the accelerations and decelerations, is not symmetric at all. Why do we say this? If two people where to give a succinct description of their experiences and these experiences where truly symmetric, one would naturally not be able to differentiate the difference in their statements because their experiences would appear exactly the same as well as their world-view via-\'a-vis the description of their experience. For example, taking the most condensed and succinct description of their experience and the swapping or interchanged some key words in their statements, the resultant statement ought to be the same as the description by the other twin. This is not the case with the present scenario.

\textbf{\underline{According to Taurai}:} He is stationery and Taurwi is moving toward Alpha-Centauri and Alpha-Centauri is not moving. 

\textbf{\underline{According to Taurwi}:} He is stationery while both the Taurai and Alpha-Centauri are moving as a whole unit like a rigid body. (Taurai and Alpha-Centauri are stationery relative to each other.) 

The description of events by the Taurai and Taurwi is not the same hence not symmetric. In order to understand what we mean here by:

\textsl{\textbf{``The description of events by each of the observers must be the same or symmetric.''}}

The reader will have to wait until the end of the next section. The asymmetry seen in the description of events here is all one needs in order to come to the conclusion that the astronaut twin Taurwi, is older when the twins reunite. What we need if we are to have a real paradox is to bring about a perfect symmetry into the whole situation. 

\subsection{\subsectionfont Twin Paradox (Symmetric)}

We shall set forth a new version of the twin paradox which is truly symmetric and this will introduce a true paradox and we shall provide a solution. Suppose  Taurai unlike in the previous version, decided to be adventurous too. He decides to rocket into space and travels not with his twin brother but all by himself and instead of Alpha-Centauri he travels at the same constant relativistic speed as Taurwi [this speed is measured by the Earth bound observers] to an imaginary constellation (call it Constellation Alpha-Christina) which is equidistant and directly opposite to Alpha Centauri along the line of sight joining the Earth and Alpha Centauri.

\begin{widetext}

\begin{figure}[h]
\centering
\epsfysize=9.8cm
\epsfbox{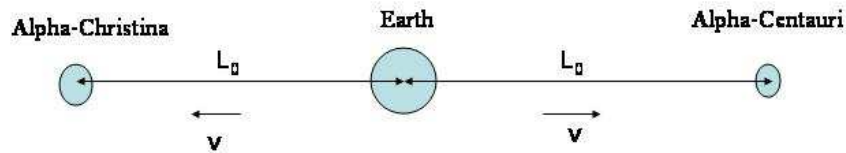}
\caption{\small \figtextfont The \figtextfont pictorial view of the symmetric twin paradox. Taurwi rockets to Alpha-Centauri at speed $V$ relative to the Earth bound observers and Taurwi rockets to the imaginary constellation Alpha-Christina which is a replica of Alpha-Centauri, at speed $V$ relative to the Earth bound observers.}
\end{figure}
\end{widetext}

On their day of departure, their family and friends bid them farewell and wish them a safe travel. Without much say, on the day of reunion, the family and friends [who all have studied physics at university and understand very well the STR] have no doubt that they [the Twins] will all have aged the same. The big question is, will the twins agree with their family and friends that they have aged the same? The truth is that, each of the twins will see the other as having aged less than they so they would not agree with their family and friends that they must be the same age. Herein we have a paradox! Who is older than who here?

If $V$ is the speed with which the Earth bound observers (family and friends) see the twins travel at, then, according to the twins in their own respective frames of references, the Earth is receding at a speed $V$ and the other twin is receding from them at a speed $2V$. This scenario is perfectly symmetric and each of the twins has every right according to the STR to say the other twin is the one that is younger and they will not agree that their ages are equal upon reuniting. We are here presented with a true paradox which the STR in its presently understood form [as is found in most if not all the textbooks of physics that deal with the subject of the STR], is unable to provide an answer. It will not help to call the GTR to our rescue because what we shall do with it for one twin, we shall have to do exactly the same with the other twin -- we have a catch here. The situation exhibits a perfect symmetry that runders us imobile as along as we stick with the provinces of Einstein's Philosophy \textit{of} Relativity.

Now, what we meant in the previous section by ``The description of event by both observers must be the same if their experience are symmetric'' is as follows:

\textbf{\underline{According to Taurai}:} He is stationery and Taurwi is receding from him at a speed $2V$ and the Earth is receding from him at a speed $V$. Alpha-Centauri is receding at a speed $V$ while Alpha-Christina is approaching him at a speed $V$. 

\textbf{\underline{According to Taurwi}:} He is stationery and Taurai is receding from him at a speed $2V$ and the Earth is receding from him at a speed $V$. Alpha-Christina is receding at a speed $V$ while Alpha-Centauri is approaching him at a speed $V$.

The above description is congruent. We just have to swap the Alpha-Christina with Alpha-Centauri and Taurai with Taurwi, that is where there is Alpha-Centauri $\longrightarrow$ Alpha-Christina where there is Alpha-Christina we make the replacement Alpha-Christina $\longrightarrow$ Alpha-Centauri and where there is Taurai $\longrightarrow$ Taurwi. It is not possible to do the same in the case of the asymmetric twin paradox of the previous section.

Once again before leaving this section, we shall re-emphasis that unlike the asymmetric twin paradox, where one can seek refuge by invoking the GTR to deal with the accelerations of one of the twins, here, this clearly won't work since both twins will all undergo the same experience. Their ages will be less than that recorded by the earth observers and these observers will measure these ages to be exactly the same but according to the twins, their ages can not be the same, hence the dilemma! How do we solve this?\\

\section{\sectionfont Proposed Solution}

The solution to the symmetric twin paradox will require us to rethink the very nimbus of the STR's central philosophy, namely that it is impossible for an inertial observer to detect their state of motion. This revision, will not alter the mathematical content of the STR, but will bring us back to the long obsoloted idea of the existence of the all pervading and permeating medium, the Aether.

Suppose we have an inertial observer $\textrm{O}$ stationed at point $\textrm{O}$  in a closed rectangular cabin OABCDE as shown in figure 2. The axis $X$ and $Y$ are orthogonal and the corners of the cabin $\perp\textrm{ABC}$, $\perp\textrm{BCD}$, $\perp\textrm{DEO}$ and $\perp\textrm{EOA}$  are right angles. At point A, observer $\textrm{O}$  places a photon emitter that emits a single photon at a time in the vertical direction parallel to EO and BC. Point D is vertically and directly above point A. Since the point D is directly above point A and the cabin OABCDE is an inertial system, according to our current understanding of inertial systems, it goes without saying that the photon emitted in the vertical direction  at point A will reach point D since light travels in straight lines. At this point D, observer $\textrm{O}$ places a photon detector that is linked to the photon emitter at point A such that observer $\textrm{O}$ is able to determine the time taken by this photon to travel from point A to point D. If $\textrm{OE=BC}=W$, the time of travel ($\Delta t$) according to observer $\textrm{O}$ of the photon will be $\Delta t=W/c$ where $c$ is the speed of light. So far so good and no problem. Lets proceed!

Let us introduce another inertial observer $\textrm{O}^{\prime}$ stationed at point $\textrm{O}^{'}$  in a closed rectangular cabin $\textrm{O}^{\prime}\textrm{A}^{\prime}\textrm{B}^{\prime}\textrm{C}^{\prime}\textrm{D}^{\prime}\textrm{E}^{\prime}$ as shown in figure 3. As is the case with the $X,Y$ axis, the axis $X^{\prime}$ and $Y^{\prime}$ are orthogonal and the corners of the cabin $\perp\textrm{A}^{\prime}\textrm{B}^{\prime}\textrm{C}^{\prime}$, $\perp\textrm{B}\textrm{C}\textrm{D}$, $\perp\textrm{D}^{\prime}\textrm{E}^{\prime}\textrm{O}^{\prime}$ and $\perp\textrm{E}^{\prime}\textrm{O}^{\prime}\textrm{A}^{\prime}$  are right angles. At point $\textrm{A}^{\prime}$, observer $\textrm{O}^{\prime}$   bores a large enough hole so much that for a photon entering via this hole, diffraction effects can be neglected and the photon can be treated as a particle.  Point $\textrm{D}^{\prime}$ is vertically and directly above point $\textrm{A}^{\prime}$. The roof of the cabin $\textrm{C}^{\prime}\textrm{D}^{\prime}\textrm{E}^{\prime}$ is photo-sensitive. Let this cabin move along the positive $x-axis$ at speed $V$ such that when the lines $\textrm{A}^{\prime}\textrm{D}^{\prime}$ and $\textrm{A}\textrm{D}$ are coincident, the photon realized at point $\textrm{A}$ by observer $\textrm{O}$ will be at the opening of the cabin $\textrm{O}^{\prime}\textrm{A}^{\prime}\textrm{B}^{\prime}\textrm{C}^{\prime}\textrm{D}^{\prime}\textrm{E}^{\prime}$ at point $\textrm{A}^{\prime}$.  So far every this looks good, lets proceed.

We have agreed that the photon can be treated here as a particle because the opening at point $\textrm{A}^{\prime}$ is large enough for us to neglect completely any diffraction effects. This photon entering at this opening will have its direction of motion being parallel to the walls, $\textrm{O}\textrm{E}$ \& $\textrm{D}\textrm{C}$ and  $\textrm{O}^{\prime}\textrm{E}^{\prime}$ \& $\textrm{D}^{\prime}\textrm{C}^{\prime}$ of the both cabins $\textrm{O}\textrm{A}\textrm{B}\textrm{C}\textrm{D}\textrm{E}$ and $\textrm{O}^{\prime}\textrm{A}^{\prime}\textrm{B}^{\prime}\textrm{C}^{\prime}\textrm{D}^{\prime}\textrm{E}^{\prime}$ respectively. Now our trouble begins!

\begin{widetext}

\begin{figure}[h]
\centering
\centering
\epsfysize=8.0cm
\epsfbox{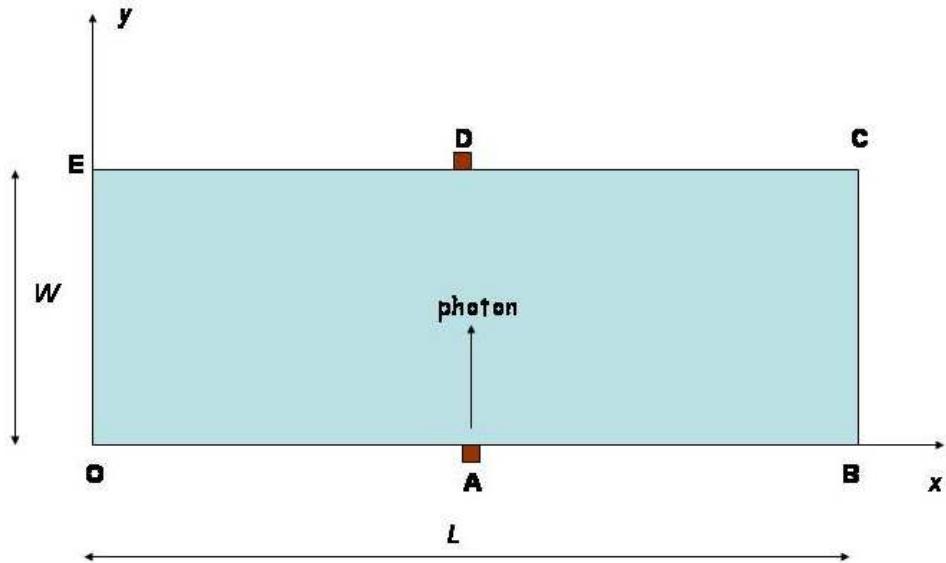}\label{a}
\caption{\small \figtextfont The closed rectangular cabin OABCDE is an inertial reference frame in which observer O is stationed at point O. Observer O has no knowledge of what is happening outside her/his cabin. S/he sends a photon vertically upwards from point A. Since light travels in straight lines, this photon is expected to reach the detector at point D.}
\end{figure}

\begin{figure}[h]
\centering
\epsfysize=8.0cm
\epsfbox{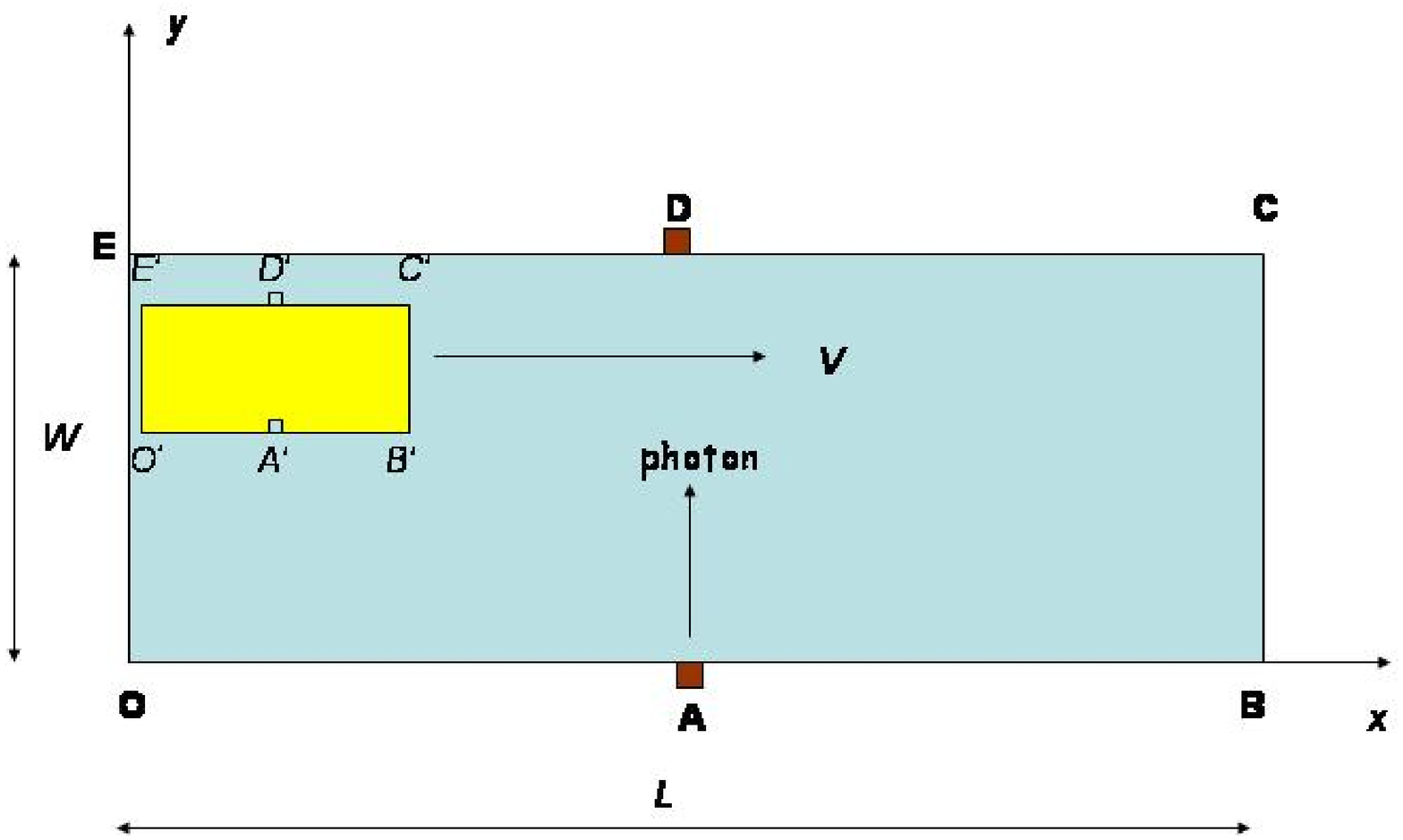}\label{b}
\caption{\small \figtextfont Inside the closed rectangular cabin OABCDE which is an inertial reference frame, we have another rectangular cabin $\textrm{O}^{\prime}\textrm{A}^{\prime}\textrm{B}^{\prime}\textrm{C}^{\prime}\textrm{D}\textrm{E}^{\prime}$ which off-cause is smaller in size compared to $\textrm{OABCDE}$. The floors and roofs of these cabins are parallel to one another respectly. In this cabin $\textrm{O}^{\prime}\textrm{A}^{\prime}\textrm{B}^{\prime}\textrm{C}^{\prime}\textrm{D}\textrm{E}^{\prime}$, we have observer $\textrm{O}^{\prime}$ stationed at point $\textrm{O}^{\prime}$. The cabin $\textrm{O}^{\prime}\textrm{A}^{\prime}\textrm{B}^{\prime}\textrm{C}^{\prime}\textrm{D}\textrm{E}^{\prime}$ moves as seen by observer $\textrm{O}$ at speed $V$ in the direction of the positive $x-axis$. The speed $V$ is such that when observer $\textrm{O}$ releases the photon from point $\textrm{A}$, this photon will reach the basement of observer $\textrm{O}^{\prime}$'s moving cabin  at point $\textrm{A}^{\prime}$.}
\end{figure}

\begin{figure}[h]
\centering
\epsfysize=8.0cm
\epsfbox{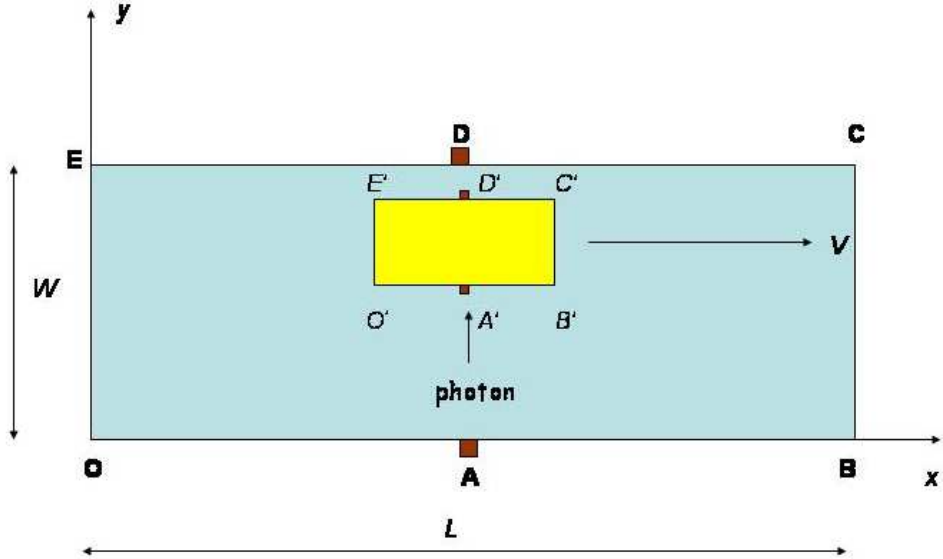}\label{c}
\caption{\small \figtextfont Just at the time when point $\textrm{A}^{\prime}$ is directly above point $\textrm{A}$, the photon released by observer $\textrm{O}$ at point $\textrm{A}$ reaches the opening at point $\textrm{A}^{\prime}$ thus enters the cabin of observer $\textrm{O}^{\prime}$. Since light travels in a straight line, will this photon continue to travel along the same straight path as seen by observer  $\textrm{O}^{\prime}$ as in figure (2) or it will travel a straight path according $\textrm{O}^{\prime}$'s cabin and reference frame?}
\end{figure}

\begin{figure}[h]
\centering
\epsfysize=8.0cm
\epsfbox{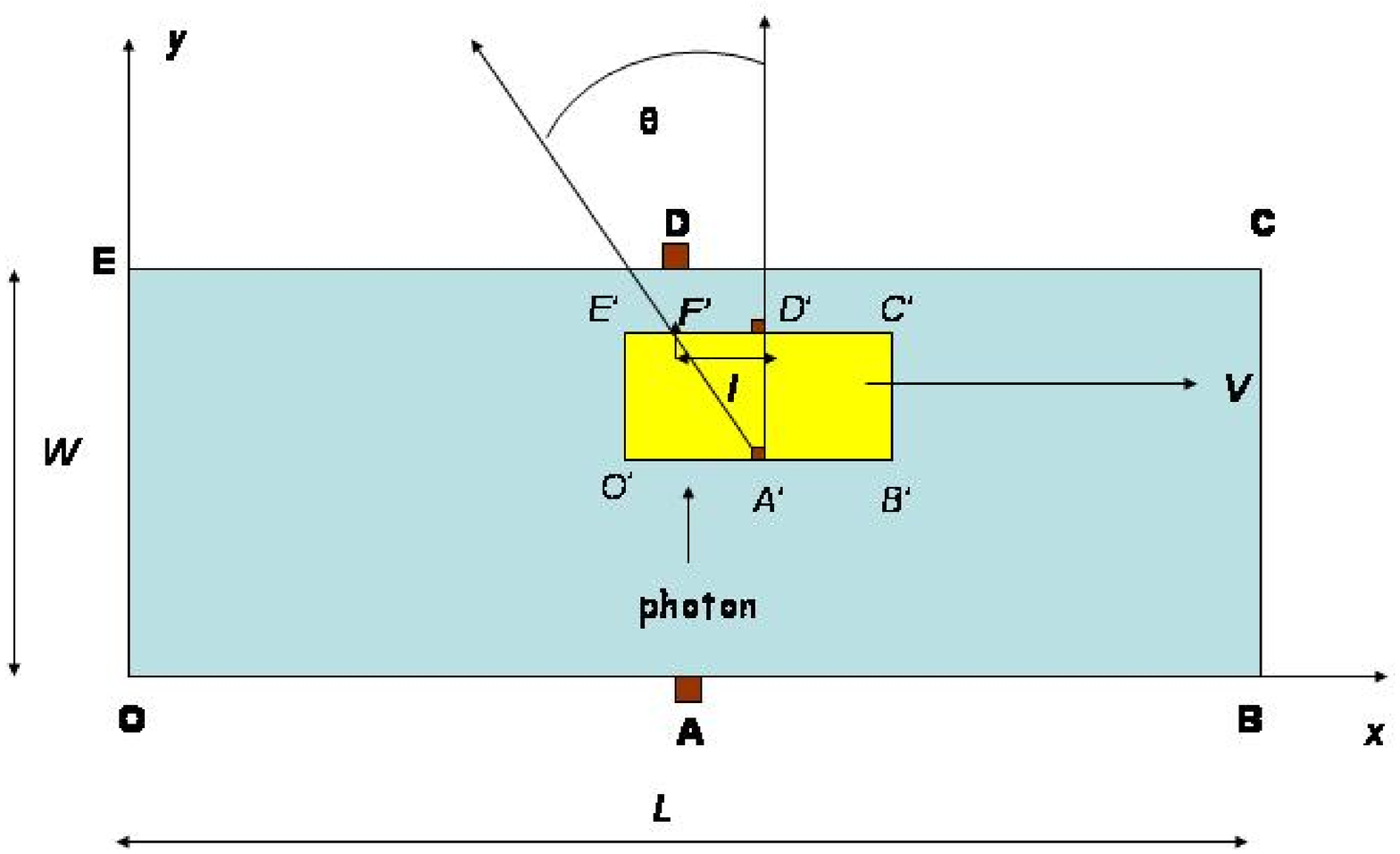}\label{d}
\caption{\small \figtextfont If the photon travels the same path as that in figure (2), then, according to observer $\textrm{O}^{\prime}$, the path of the photon will be inclined at an angle $\theta$ to her/his walls and this photon will traverse path $\textrm{A}^{\prime}\textrm{F}^{\prime}$ and not $\textrm{A}^{\prime}\textrm{D}^{\prime}$ as would expected for a photon traveling in the vertical direction from point $\textrm{A}^{\prime}$ in the cabin $\textrm{O}^{\prime}\textrm{A}^{\prime}\textrm{B}^{\prime}\textrm{C}^{\prime}\textrm{D}\textrm{E}^{\prime}$. If the photon traversed along a straight path according $\textrm{O}^{\prime}$ as much as in the case figure (2), then the photon will have to exit the cabin $\textrm{O}^{\prime}\textrm{A}^{\prime}\textrm{B}^{\prime}\textrm{C}^{\prime}\textrm{D}\textrm{E}^{\prime}$ at point  $\textrm{D}^{\prime}$ which will be offset from the point $\textrm{D}$ when it reaches the roof of $\textrm{O}$ 's cabin;  thus the photon will have to be detected by observer $\textrm{O}$ at a point off-set from the initial point $\textrm{D}$ and this new point will be to the right-side of point $\textrm{D}$. If this is the case, observer $\textrm{O}$  is forced to draw the conclusion that the cabin $\textrm{O}^{\prime}\textrm{A}^{\prime}\textrm{B}^{\prime}\textrm{C}^{\prime}\textrm{D}\textrm{E}^{\prime}$ at point affected the motion of the photon.}
\end{figure}
\end{widetext}

Since $\textrm{O}^{\prime}$ is a inertial observer and s/he has knowledge that the particle that just entered is a photon and the direction of motion of this photon is as aforedescribed. The question is; Will s/he see the photon continue to travel parallel to the walls of her/his cabin? If it does, then, s/he will expect at some finite time in the future that this photon will be detected at point  $\textrm{D}^{\prime}$. If it so happens that at this point $\textrm{D}^{\prime}$, we have an opening, the photon will travel outside the cabin of observer $\textrm{O}^{\prime}$ upon arriving at point $\textrm{D}^{\prime}$ and this photon will be detected on the roof of observer  $\textrm{O}$'s cabin albeit off-set from point $\textrm{D}$ (to the right-side of). The reason the photon will be detected off-set the point $\textrm{D}$ is because at the time of exit of the photon at point $\textrm{D}^{\prime}$, this point is no-longer directly above point $\textrm{D}$ because this cabin is moving relative to the cabin of observer $\textrm{O}$ and the photon will have to continue its journey in a straight line parallel to wall of both cabins.

Let us re-state or rephrase what we have just said in the previous paragraph. If the photon travels the same path as that in figure ($2$), then, according to observer $\textrm{O}^{\prime}$, its path will be inclined at an angle $\theta$ to her/his walls and this photon will traverse path $\textrm{A}^{\prime}\textrm{F}^{\prime}$ and not $\textrm{A}^{\prime}\textrm{D}^{\prime}$ as would expected for a photon traveling in the vertical direction from point $\textrm{A}^{\prime}$ in the cabin $\textrm{O}^{\prime}\textrm{A}^{\prime}\textrm{B}^{\prime}\textrm{C}^{\prime}\textrm{D}\textrm{E}^{\prime}$. This angle $\theta$ is such that:

\begin{equation}
V=c\tan\theta.\label{v}
\end{equation}

If the photon traversed along a straight path according $\textrm{O}^{\prime}$ just as in the case figure ($2$), then the photon will have to exit the cabin $\textrm{O}^{\prime}\textrm{A}^{\prime}\textrm{B}^{\prime}\textrm{C}^{\prime}\textrm{D}\textrm{E}^{\prime}$ at point  $\textrm{D}^{\prime}$ which will be offset thus the photon will have to be detected by observer $\textrm{O}$ off-set from the point $\textrm{D}$ to the right-side. If this is the case, observer $\textrm{O}$  is forced to draw the conclusion that the cabin $\textrm{O}^{\prime}\textrm{A}^{\prime}\textrm{B}^{\prime}\textrm{C}^{\prime}\textrm{D}\textrm{E}^{\prime}$ at point  $\textrm{D}^{\prime}$ affected the motion of the photon. At this point we leave the reader to decide whether they think the photon will exhit from $\textrm{O}^{\prime}$ at point $\textrm{D}^{\prime}$ or not. We believe the reader will reach the conclusion that the photon will not exit via point $\textrm{D}^{\prime}$ but would exit at point $\textrm{F}^{\prime}$ is a hole was bored there and that observer $\textrm{O}$ will then be able to detect the photon at point $\textrm{D}$. Accepting this, amounts to accepting that observer $\textrm{O}^{\prime}$ will be able to determine his speed in-accordance with equation (\ref{v}).

From the foregoing, the solution to the symmetric twin-paradox is clear. Taurai and Taurwi can determine their state of motion and even measure their velocity which they will each find to be $V$ and this velocity is their velocity relative to some absolute and universal medium that is absolute rest and this medium clearly must be the one in which light has this constant speed $c$. If the Laws \textit{of} Nature are to be the same everywhere in space and time, then, it follows that this medium must fill all of space. The length contraction and time dilation occur relative to this medium and these properties are exactly as those of the Lorentz Aether. We are thus are brought back to the old ideas that now ``safely'' belongs to the Science Museum \textit{of} Great \textit{but} Failed Ideas.

\section{\sectionfont Discusion and Conclusions}

The idea of a universal all-pervading and permeating medium, is a superseded idea. Einstein's 1905 STR rendered it obsolete and ever since then, research on this idea is not taken seriously hence the reason for the lack of citation of recent reseach on this field. The more than century old MM-Exp is said to be enough proof against this idea and it is said/thought/and or supposed, that this experiment alone closed down the curtain once and for all on this subject. 

If the arguments presented in this reading are correct, then, we are called back to the drawing board to rethink our long held belief that a universal all-pervading and permeating medium, is a superseded idea. This believe stems from the fact the STR proclaimed that it is impossible for an inertial observer to detect their state of motion.  We have shown here that not only is an observer able to determine where or not they are moving, but that they are able to deduce their velocity. This velocity will have to measured relative to the same ``medium'' in which light is the speed of light has this same speed $c$.

We realize, that if our arguments are correct, then, not only is the speed of light the same for all inertial observers, but the direction of motion of this light. This would mean we have to re-write the second postulate of the STR which in most physics texts book reads (see e.g Cutnell \& Johnson 2003; Halliday \& Resnick Walker 1997):

\begin{quote}
\textsl{``The speed of light in vacuum has the same value $c$ in all directions and in all inertial reference frames"}
\end{quote}

to read:

\begin{quote}
\textsl{``The velocity of light in a gravity free vacuum is the same for all observers."}\\
\end{quote}

The term ``velocity'' is different from ``speed'' as this term [velocity] includes the speed and the direction of propagation of the beam of light. What this means is that all inertial observers will agree not just on the speed but on the direction of propagation of the beam. In the case as presented in figures (2)-(5), the photon will not change its direction of motion relative to observer $\textrm{O}$ thus the meaning of will be that, $\textrm{O}^{\prime}$ will see the photon traverse at an inclined angle $\theta$ to her/his walls. We have already argued, this angle is enough to deduce the speed of the cabin.

Once again, if these ideas are correct, what this really means  is that time dilation and length contraction are real physical phenomena in much the same way as Lorentz (1892, 1904) and Fitz-Gerald (1889)  envisaged and the Aether is also real in the sense envisaged by Maxwell (1973) (and many advocated it the Aether theory) when he propounded his electromagnetic theory which Einstein mused until he arrived at the STR.

The shift $\Delta l$ measured by observer $\textrm{O}^{\prime}$ in his/her cabin as shown in figure (5) is given:

\begin{equation}
\Delta l=\left(\frac{V}{c}\right)W^{\prime}\label{shift}
\end{equation}

where $W^{\prime}$ is the height of the cabin. This can be generalized to any given inertial observer. Thus if one is in an inertial frame of reference and they projected a light beam vertically up-wards and this light beam strikes the roof not on a point directly above the point when the beam of light was realized, the conclusion they have to make is that their cabin of system of reference is in motion and the shift is related to the speed of their frame of reference and the height of this system by equation (\ref{shift}). 

Given that the gravitational pull between the Earth and Sun causes the Earth to travel around, or orbit, the Sun at a speed of $30\, \textrm{kms}^{-1}$ this would mean a laboratory that is say $10\,\textrm{m}$ in height will be expected to register a shift in accordance with equation (\ref{shift}), of about $100\mu\textrm{m}$. Given modern day precision, it should be possible to detect such a shift. Therefore, we propose that this experiment be carried out. Should the results provide a negative result, the present ideas are immediately rendered null and void and Einstein's philosophy about the obsoleteness of absolute motion is holds. If the experiments prove this shift, then, nothing of the mathematical structure of the STR will change, expect its philosophy and this philosophy will exactly be that championed by Lorentz in his works (Lorentz 1892, 1904).\\
\\

\end{document}